# Social Network Based Search for Experts

Yehonatan Bitton
Dept. Info. Systems Engineering
Ben-Gurion University of the Negev
Beer-Sheva, 84105, Israel
lolporer@gmail.com

Michael Fire
Telekom Innovation Lab & Dept. Info. Systems Engineering
Ben-Gurion University of the Negev
Beer-Sheva, 84105, Israel
mickyfi@gmail.com

Dima Kagan
Dept. Info. Systems Engineering
Ben-Gurion University of the Negev
Beer-Sheva, 84105, Israel
dimakagan15@gmail.com

Bracha Shapira
Dept. Info. Systems Engineering
Ben-Gurion University of the Negev
Beer-Sheva, 84105, Israel
bshapira@bgu.ac.il

Lior Rokach
Dept. Info. Systems Engineering
Ben-Gurion University of the Negev
Beer-Sheva, 84105, Israel
liorrk@bgu.ac.il

Judit Bar-Ilan
Dept. Information Science
Bar-Ilan University
Ramat Gan, 52900, Israel
Judit.Bar-Ilan@biu.ac.il

## ABSTRACT
Our system illustrates how information retrieved from social networks can be used for suggesting experts for specific tasks. The system is designed to facilitate the task of finding the appropriate person(s) for a job, as a conference committee member, an advisor, etc. This short description will demonstrate how the system works in the context of the HCIR2012 published tasks.

## Categories and Subject Descriptors
H.3.3 [**Information Storage and Retrieval**]: Information Search and Retrieval – *Selection process.*

## General Terms
Algorithms, Design, Experimentation,

## Keywords
Social networks, Text categorization, HCIR.

## 1. INTRODUCTION
Social networks have become very popular, not only in everyday life (e.g. Facebook or Google+), but in the academic setting as well (e.g. Academia.edu[1] or ResearchGate[2]). Online reference managers like Mendeley[3] or CiteULike[4] also have a social network component – groups of people with common interests can be formed. In addition implicit social networks can be constructed based on the information available in the system, such as common interests of the registered users, and coauthorship of the papers "bookmarked" by the users of the system.

It is widely accepted that the number of citations an article or an author receive implies "impact" [7]. Recently, new indicators of impact have been suggested [8], and these include reader counts on reference managers. It was shown that there is medium-high and significant correlation between citation counts and readership counts [1-2,6]. Correlations are high but far from perfect, indicating that readership counts available from reference managers like Mendeley convey an impact that is not identical to citation-based indices. Correlation was also found between leadership roles inside an organization's social network and different centrality measures, such as closeness centrality and PageRank [3].

For this challenge we were provided with a partial dataset from Mendeley, that includes the profiles of more than one million Mendeley users, about 145,000 publications that were bookmarked by Mendeley users (incl. the number of readers of each item, and the distribution of the disciplines and academic statuses of the readers), social network of the users and the members of public groups. We enriched the Mendeley datasets by cross-referencing the Mendeley users' profiles with 744,884 users' profiles which were collected from Academia.edu by dedicated web crawler [4].

The example tasks in the challenge included finding the right person for a job, recruiting PC members for a conference and to provide name of experts that can provide some testimony. We adopted a social network representation of the available data. For ranking the output we utilized the "wisdom of crowds" based on readership counts.

## 2. THE TASKS
Three example tasks were provided by the challenge's organizers. In section 5, we will describe and demonstrate how the system accomplishes the second task:

1. **Hiring**
   Given a job description, produce a set of suitable candidates for the position. An example of a job description: http://www.linkedin.com/jobs?viewJob=&jobId=3004979.

2. **Assembling a Conference Program**
   Given a conference's past history, produce a set of suitable candidates for keynotes, program committee members, etc. for the conference. An example conference could be HCIR 2013, where past conferences are described at http://hcir.info/.

---
[1] http://academia.edu/

[2] http://www.researchgate.net

[3] http://www.mendeley.com

[4] http://www.citeulike.org

3. **Finding People to deliver Patent Research or Expert Testimony**

   Given a patent, produce a set of suitable candidates who could deliver relevant research or expert testimony for use in a trial. These people can be further segmented, e.g., students and other practitioners might be good at the research, while more senior experts might be more credible in high-stakes litigation. An example task would be to find people for http://www.articleonepartners.com/study/index/1658-system-and-method-for-providing-consumer-rewards.

## 3. THE SEARCH INTERFACE

Our system is designed for easy user interaction (see Figure 1). It has a text area (1 in Figure 1), a Search button (2) an "I'm Feeling Lucky" button and four checkboxes (4) that allow to limit the names of the experts retrieved only to those with the chosen academic status(es). This interface is available at http://proj.ise.bgu.ac.il/15/ExpertRecommendation/index.html.

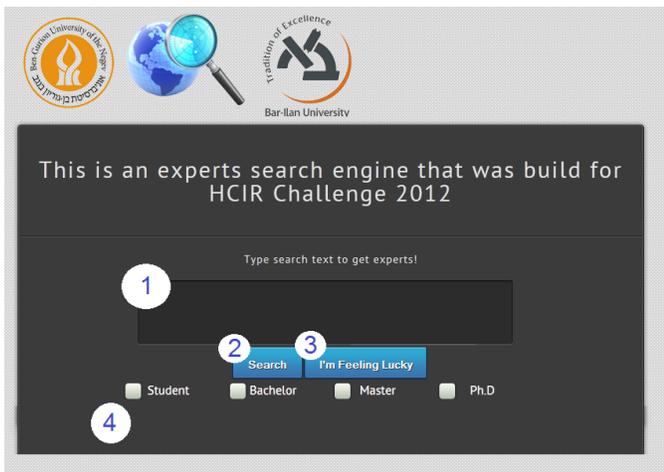

Figure 1: Home page and search interface of our system

## 4. BEHIND THE INTERFACE

For the first step, the system uses the Yahoo API for categorizing the text that is given in the UI. The Yahoo API returns several results ordered from the most fitting one to the less. When the keywords are sent to the server we are using Levenshtein distance [5] to compute distance between the keyword that was sent and the words from Mendeley categories the category with the lower distance is the most fitting key word for the search task.

The heart of the system is based on creating a social network over the people who are in the database both as having profiles and authors from the publication that are in the provided dataset (Figure 2).

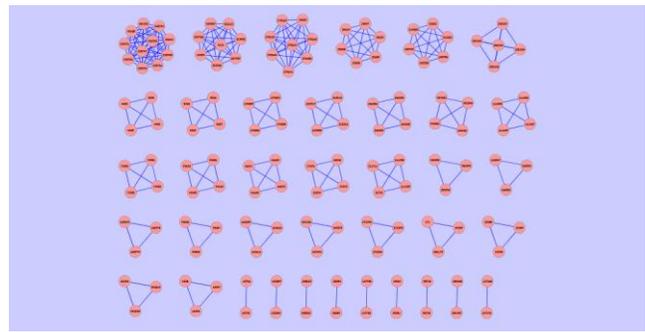

Figure 2 the coauthors network for the authors only.

In Figure 2 we can see that due to sparseness of the data we have a high number of cliques which makes the task much more difficult. First, we have used Microsoft Academic Search in-order to categorize each publication and journal, this way we can decide whether a person wrote a publication in a journal which helps us creating the social network.

The next step was to add the profiles social network which is more complete than the coauthorships network and helped us determine who is really an expert in the system (Figure 3).

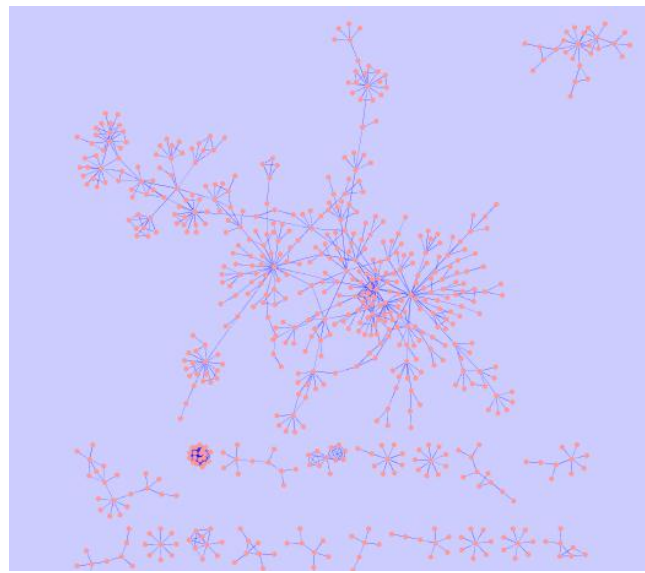

Figure 3: The complete social network.

We are extracting the following features for each person: Page Rank, betweenness and closeness centrality, journals ranks, number of readers, all based on the weight of each edge, where each edge weight is determined by the number of publications coauthored and the connection from the profiles network of Mendeley.

Another feature which help us decide which person is most fitting to this category is the user rank feature, each user can add 1 or subtract 1 to the person's rank.

We have created a module that learns from a dataset the appropriate features rates and this is how we decide in the net who are the experts in the specific social network of the domain. The learning was accomplished using C4.5 decision tree [9].

Another problem that we had was matching between authors and profiles this was done using the same technique as before the Leveshtein distance , we have computed each author distance and the one with lowest was matched to the profile. If there was more than one equal match we have discarded the connection.

## 5. DEMONSTRATING TASK No. 2

We start with entering the text of task no. 2 to the text box of the search interface (see Figure 4). The academic status of the required person can also be chosen.

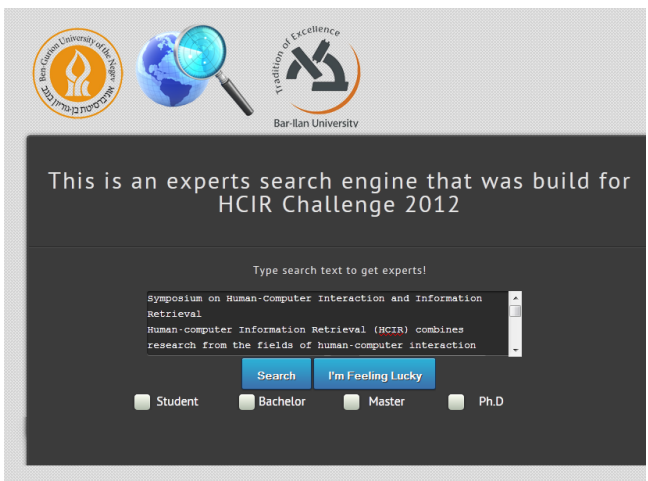

Figure 4: The text of task no. in the search box

Now the system suggests several content categories, as shown in Figure 5.

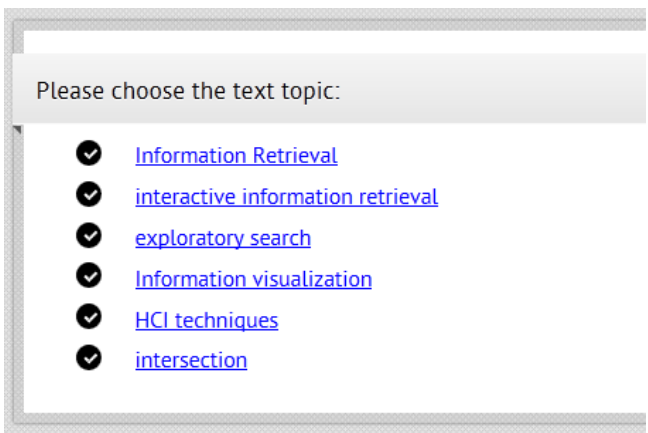

Figure 5: Content categories based on the input

After choosing the best-fitting category (information retrieval in this example), a ranked list of names is produced as can be seen in Figure 6.

Each person's name is linked to a page which have the person research interests and publications, moreover as you can see there are two icons by each name for improving the system: if a user clicks the red X this is interpreted as a vote against the current rank of the person, and the green checkmark is vote for the person.

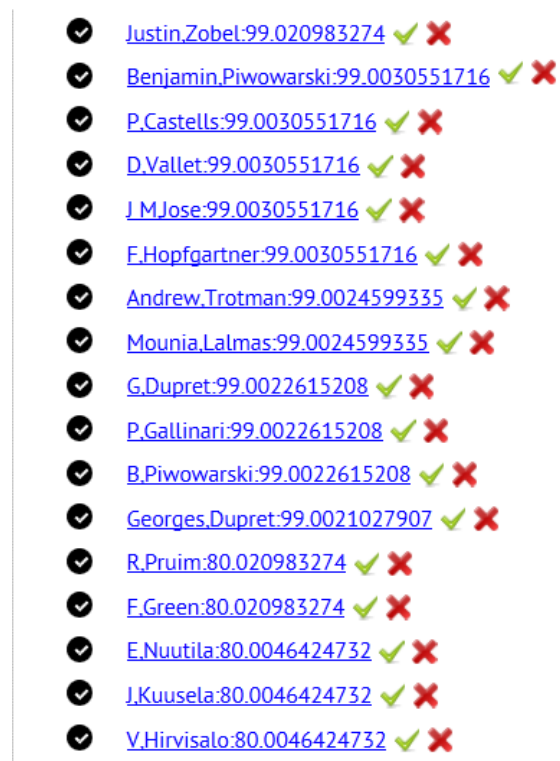

Figure 6: The ranked list of experts in the specific area

## 6. CONCLUSIONS

The system presented here shows how social network analysis and text categorization can be utilized for identifying experts. The system is based on the partial data received from Mendeley, and as such its performance in not optimal. There are several possibilities for enhancing the system in the future.